\documentclass[aps,pra,twocolumn,showpacs,groupedaddress]{revtex4}
\usepackage{graphicx}

\newcommand{\ket}[1]{|#1\rangle}
\newcommand{\bra}[1]{\langle#1|}

\newcommand{\schrodinger}{Schr\"{o}dinger}

\newcommand{\sigmam}{\sigma^-}
\newcommand{\sigmap}{\sigma^+}
\newcommand{\inner}[2]{\langle #1|#2\rangle} %ÄÚ»ý
\newcommand{\p}[1]{\frac{\partial}{\partial t}#1} %¶ÔtµÄÆ«µ¼Êý

\newcommand{\PRA}[3] {Phys. Rev. A {\bf #1}, #2
(#3)}
\newcommand{\PRD}[3] {Phys. Rev. D {\bf #1}, #2
(#3)}
\newcommand{\PRL}[3] {Phys. Rev. Lett. {\bf #1}, #2
(#3)}

\newcommand{\JPB}[3] {J. Phys. B {\bf #1}, #2 (#3)}
\newcommand{\PLA}[3] {Phys. Lett. A {\bf #1}, #2 (#3)}
\newcommand{\EPL}[3] {Europhysics Letters {\bf #1}, #2 (#3)}

\newcommand{\PS}[3] {Phys. Scr. {\bf #1} #2
(#3)}

\begin{document}
%\begin{CJK*}{GBK}{song}
%\draft
\title{Born-Oppenheimer approximation in open systems}
\author{X. L. Huang}%\footnote{ghost820521@163.com} }
\author{X. X. Yi\footnote{yixx@dlut.edu.cn}}

\affiliation{ School of Physics and Optoelectronic Technology,\\
Dalian University of Technology, Dalian 116024 China}

\date{\today}

\begin{abstract}
We generalize the standard Born-Oppenheimer approximation  to the
case of open quantum systems. We define the zeroth order
Born-Oppenheimer approximation  for  an open quantum system as the
regime in which its effective Hamiltonian can be diagonalized with
fixed slowly changing variables. We then establish validity and
invalidity conditions for this approximation for two types of
dissipations---the spin relaxation and the dissipation of
center-of-mass motion. As an example, the Born-Oppenheimer
approximation  for a two-level open system is analyzed.
\end{abstract}

 \maketitle
%\end{CJK*}

\section{Introduction}
\schrodinger~equation which governs the dynamics of quantum system
is among the  fundamental equations in  physics. In principle we can
predict all features  of a  quantum system at zero temperature by
solving this equation. However, many problems are hard to handle and
the \schrodinger~equation might not be solved analytically. Many
approximations have been developed in order to solve this problem,
including the adiabatic approximation \cite{adiabaticBorn1928} and
the Born-Oppenheimer(BO) approximation  \cite{BO1927}.

The adiabatic approximation tells us that for a sufficiently slowly
varying Hamiltonian, if the system initially is in an instantaneous
eigenstate of the Hamiltonian,  the system  will remain in that
eigenstate up to a phase factor at later times (for the adiabatic
condition and recent progress, see
Ref.\cite{adiabaticMacKenzie2007,adiabaticWu2005,adiabaticMarzlin2005,
adiabaticTong2005,adiabaticTong2007,adiabaticYe2007,adiabaticZhao2008,
adiabaticDu2008,adiabaticFujikawa2008,adiabaticComparat2009}). When
the quantum system evolves adiabatically and cyclically, the
eigenstate of the Hamiltonian will acquire a Berry phase
\cite{Berry1984} except the dynamical phase. This concept was
generalized to non-cyclic and non-adiabatic cases, and the
non-dynamical phase acquired in this situation was called the
geometric phase
\cite{GPRezakhani2006,GPSjoqvist2000,GPAA1987,GPHuang2008,
GPMoller2008,GPTong2004,GPWang2007,GPYang2007}. If a quantum system
has two sets of variables, say a fast  and a slow set of variables,
the BO approximation can be used to solve the dynamics, this can be
done  by solving the fast variables with fixed slow variables first,
leading to an effective Hamiltonian for  the slow variables.
Applying this approximation to a bipartite particle, an external
vector potential is  introduced for the slowly moving particle due
to its fast moving partner \cite{CPSun1990}. Thus the total
wave-function can be factorized to a product of two wave-functions
corresponding to the fast and the slow variables. The BO
approximation has been widely used in many fields of physics and
chemistry and proved to be a fundamental tool in these research
\cite{Cederbaum2008}.

These two approximations  are both for  closed quantum systems.
However, due to the unavoided couplings of quantum system to its
environment, a realistic quantum system is in general open and  the
dynamics of such a system can be described  by the so-call Markovian
master equation. The master equation was derived from
\schrodinger~equation by tracing out the environment variables in
the Markovian limit \cite{quantumnoisy} (for the solution of this
equation, see e.g.
Ref.\cite{solutionLiemethod2003,solutionKrausmethod2006,solution1,
solution2,solutiondampingbasis1993,solutionperturbative2000} and for
the dynamics beyond the Markovian limitation, see
Ref.\cite{nMarkovBreuer2006,nMarkovBreuer2007,nMarkovHuang2008,
nMarkovManiscalco2006,nMarkovManiscalco2007,nMarkovShabani2005,
nMarkovVacchini2008,nMorkovbook2007,nMarkovBreuer2008}). The
adiabatic approximation has been extended to the case of open
systems by several authors
\cite{EHAYi2007,EHAHuang2008,AOTong2007,AOThunstrom2005,AOSarandy2005,
AOHara2008,AOFleischer2005}. One of these extensions was done by the
effective Hamiltonian approach
\cite{EHAHuang2008,EHAYi2007,AOTong2007}. The key idea of the
effective Hamiltonian approach can be outlined as follows. By
introducing an ancilla which is the same as the quantum system, we
can map the density matrix of the quantum  system to a wave-function
of the composite system (system plus ancilla). Then  the master
equation can be mapped to a \schrodinger-like equation. Since the
dynamics of open systems is governed by a \schrodinger-like
equation, some conclusions for the closed system can be extended to
open systems straightforwardly. In this paper, we extend the BO
approximation to open systems by using the effective Hamiltonian
approach. Two types of dissipation, the spin relaxation and the
dissipation in the center-of-mass motion(DissCOM), will be
considered. The validity condition for the BO approximation is also
presented.

The structure of this paper is organized as follow. In
Sec.\ref{sec:spindissipation} we generalize  the BO approximation to
open systems with spin relaxations. The case of dissipation in the
center-of-mass motion (DissCOM) is considered in
Sec.\ref{sec:coordinatedissipatin}. In Sec.\ref{sec:example} we
present an  example for the case of  spin relaxation to show the
details  of the formulism.  Conclusions and discussions are given in
Sec.\ref{sec:CD}.

\section{Spin relaxation\label{sec:spindissipation}}

%\subsection{Coupled basis expansion\label{sec:Coupled}}
Consider a spin moving in a magnetic field. Taking the spin
relaxation into account, the evolution of the density matrix for
such a system can be described by the Markovian master equation that
generally can be written in the following form ($\hbar=1$)
\cite{quantumnoisy},
\begin{eqnarray}
\p \rho=-i[H,\rho]+\mathcal{L}\rho,\label{masterequation}
\end{eqnarray}
where $H$ is the Hamiltonian for the system. In our discussion, this
Hamiltonian can be divided into the following three parts
\begin{eqnarray}
H={H}_K+V+H_S(\vec{B}(\vec{x}),\vec{S}),\label{Hamiltonian}
\end{eqnarray}
where ${H}_K={H}_K(\vec{x})=\frac{{\vec{p}}^2}{2M}$ is the kinetic
energy, ${\vec{p}}$ is the momentum operator, $V=V(\vec{x})$ denotes
an external  potential and $H_S(\vec{B}(\vec{x}),\vec{S})$
represents the coupling of the spin to  an external magnetic field
$\vec{B}(\vec{x})$. Note that the coupling   depends on the
coordinate $\vec{x}$ through the inhomogeneous external field
$\vec{B}(\vec{x})$. The second term on the right hand side of
Eq.(\ref{masterequation}) denotes the spin relaxation/dissipation,
which is assumed here to takes the Lindblad form \cite{Lindblad},
\begin{eqnarray}
\mathcal{L}\rho=\frac12\gamma(2\sigmam\rho\sigmap-
\rho\sigmap\sigmam-\sigmap\sigmam\rho),\label{spindissipation}
\end{eqnarray}
where $\sigmap$ and $\sigmam$ are Pauli operators. For the case of
DissCOM, this term is different \cite{quantumnoisy}. We shall
discuss  this problem  in Sec.\ref{sec:coordinatedissipatin}.

Following the standard procedure of the effective Hamiltonian
approach \cite{EHAYi2001,EHAYi2007,EHAHuang2008}, we can map the
density matrix into a wave-function by introducing an ancillary
system, and obtain a \schrodinger-like equation
\cite{EHAYi2001,EHAYi2007,EHAHuang2008},
\begin{eqnarray}
&&\rho\rightarrow\ket{\widetilde{\Phi}}=
\sum_{m,n=1}^N\rho_{mn}(t)\ket{E_m}\ket{e_n},\nonumber\\
&&i\p\ket{\widetilde{\Phi}}=\mathcal{H}_T\ket{\widetilde{\Phi}},
\end{eqnarray}
where $\rho_{mn}(t)=\bra{E_m}\rho(t)\ket{E_n}$ is the element of the
density matrix. $\{\ket{E_n}\}$ and $\{\ket{e_n}\}$ are the
time-independent  bases for the spin and the ancilla, respectively.
$\mathcal{H}_T$ is the so-called non-Hermitian effective
Hamiltonian. For the considered moving magnetically-driven
spin(Eq.(\ref{Hamiltonian})) and the master equation
(\ref{masterequation}), the effective Hamiltonian has the following
form,
\begin{eqnarray}
\mathcal
{H}_T={H}_K-{H}_K^{\text{A}}+V-V^{\text{A}}+H_S-H_S^{\text{A}}+L_S,\label{EH}
\end{eqnarray}
where the superscript $\text{A}$ denotes  operators of the auxiliary
system, which are defined by
\begin{eqnarray}
\bra{e_m}O^{\text{A}}\ket{e_n}=\bra{E_n}O^{\dag}\ket{E_m},
\end{eqnarray}
and ${H}_K^{\text{A}}={H}_K^{\text{A}}(\vec{x}^{\text{A}})$,
$V^{\text{A}}=V^{\text{A}}(\vec{x}^{\text{A}})$,
$H_S^{\text{A}}=H_S^{\text{A}}(\vec{B}(\vec{x}^{\text{A}}),\vec{S}^{\text{A}})$,
 $L_S$ represents the spin relaxation, its form  depends on the coupling
 of the  system to its environment and hence it is a function of the jump operators.
 For spin relaxation/dissipation (Eq.(\ref{spindissipation})), we
 have
 \begin{eqnarray}
L_S=i\gamma\sigmam\tau^--\frac{i}2\gamma\sigmap\sigmam-\frac{i}2\gamma\tau^+\tau^-,
\end{eqnarray}
where $\tau^+$ and $\tau^-$ are Pauli operators for the auxiliary
system. Notice  that there is only one term in $L_S$ describes  the
interaction between the quantum system and the ancilla. Now we are
in a position to extend the BO approximation from closed to open
quantum systems. In order to simplify the derivation and make the
formulation  more explicit, we define a new six-dimension coordinate
by the coordinates of  the quantum system and the ancilla as
$\vec{r}{=}(\vec{x},i\vec{x}^{\text{A}})$. With these notations, the
effective Hamiltonian can be rewritten as
$\mathcal{H}_T={H}_K^{\text{T}}(\vec{r})+V^{\text{T}}(\vec{r})+H_S^{\text{T}}(\vec{r})$
with
${H}_K^{\text{T}}(\vec{r})={H}_K(\vec{x})-{H}_K^{\text{A}}(\vec{x}^{\text{A}})$,
$V^{\text{T}}(\vec{r})=V(\vec{x})-V^{\text{A}}(\vec{x}^{\text{A}})$
and
$H_S^{\text{T}}(\vec{r})=H_S(x)-H_S^{\text{A}}(x^{\text{A}})+L_S$.
$H_S^{\text{T}}(\vec{r})$ is the total spin part of the effective
Hamiltonian $\mathcal{H}_T$. We denote  the right eigenfunctions of
$H_S^{\text{T}}(\vec{r})$ by
$\ket{\Lambda_n^R}=\ket{\Lambda_n^R(\vec{r})}$, $(n=1,2,...,N^2)$
and the left eigenfunctions by
$\bra{\Lambda_n^L}=\bra{\Lambda_n^L(\vec{r})}$ with the
corresponding eigenvalues $E_n=E_n(\vec{r})$ for a fixed $\vec{r}$.
Generally speaking, the total spin part $H_S^{\text{T}}$ is not
Hermite and the eigenvalues are usually complex. Nevertheless the
left and right eigenfunctions still satisfy the orthonormal relation
$\inner{\Lambda_m^L}{\Lambda_n^R}=\delta_{mn}$. We should note that
when we map the density matrix  into a  wave-function of the
composite system, degeneracies may be introduced. The occurrence of
degeneracies is a signature of the presence of symmetry in the
composite system \cite{Tay2007PRA}. In the following, we have
restricted our discussion on the non-degenerate energy levels.

We can expand the eigenstate of the effective Hamiltonian
$\mathcal{H}_T$ in terms of $\{\ket{\Lambda_{n}^R}\}$ as
\begin{eqnarray}
\ket{\widetilde{\Phi}}=\sum_{k=1}^{N^2}\Phi_{n}(\vec{r})\ket{\Lambda_{n}^R(\vec{r})}.
\label{expansion}
\end{eqnarray}
Substituting Eq.(\ref{expansion}) into the eigenvalue equation of
the effective Hamiltonian
$\mathcal{H}_T\ket{\widetilde{\Phi}}=E\ket{\widetilde{\Phi}}$, we
obtain the following  equation for $\Phi_{n}(\vec{r})$,
%\begin{widetext}
\begin{eqnarray}
{H}^{\text{T}}(n)\Phi_{n}+F^{\text{T}}(n)\Phi_{n}+\sum_{m\neq
n}{O}^{\text{T}}(n,m)\Phi_{m}=E\Phi_{n},\label{phieqn}
\end{eqnarray}
%\end{widetext}
where
\begin{eqnarray}
&&{H}^{\text{T}}(n){=}{-}\frac1{2M}[\nabla_{\vec{r}}{-}i\vec{A}^{\text{T}}_{\vec{r}}(n)]^2{+}V^{\text{T}}(\vec{r}){+}E_n(\vec{r}),\\
&&\vec{A}^{\text{T}}_{\vec{r}}(n){=}i\inner{\Lambda_n^L}{\nabla_{\vec{r}}\Lambda_n^R},\\
&&F^{\text{T}}(n){=}{-}\frac1{2M}\sum_{m\neq
n}\left(\inner{\Lambda_n^L}{\nabla_{\vec{r}}\Lambda_m^R}\inner{\Lambda_m^L}{\nabla_{\vec{r}}\Lambda_n^R}\right),\\
&&{O}^{\text{T}}(n,m){=}{-}\frac1{2M}\left(2\inner{\Lambda_n^L}{\nabla_{\vec{r}}\Lambda_m^R}\nabla_{\vec{r}}{+}
\inner{\Lambda_n^L}{\nabla^2_{\vec{r}}\Lambda_m^R}\right).~~~~~
\end{eqnarray}
$\nabla_{\vec{r}}$ in the above equations denotes the derivative
with respect to the new coordinates $\vec{r}$. We can see from these
equations  that $\ket{\nabla_{\vec{r}}\Lambda_m^R}=0$  for all $m$
may lead  $F^{\text{T}}$ and ${O}^{\text{T}}$ to vanishing, this
results in a complete separation of the spin and motion variables in
the dynamics, which  is quite different from closed systems. It is
easy to find one solution that these condition can be met when the
dissipation is negligible and the external field is homogeneous.
Treat $F^{\text{T}}$ and ${O}^{\text{T}}$ as perturbations, we can
use the standard perturbation theory to solve Eq.(\ref{phieqn}).
First we rewrite Eq.(\ref{phieqn})  as
\begin{eqnarray}
(\mathcal{H}_0+\mathcal{W})\Phi=E\Phi \label{phieqnmatrixform}
\end{eqnarray}
where $\Phi$, $\mathcal{H}_0$ and the perturbation $\mathcal{W}$ are
defined as
\begin{eqnarray*}
\begin{array}{c}
\Phi=\left[
\begin{array}{c}
  \Phi_{1}  \\
  \Phi_{2} \\
  \vdots\\
  \Phi_{N^2}\\
  \end{array}
\right],
\end{array}
\begin{array}{c}
\mathcal{H}_0=\left[%
\begin{array}{cccc}
   {H}^{\text{T}}(1)&\  0  &\ \cdots  &\ 0 \\
  0 &\  {H}^{\text{T}}(2) &\ \cdots   &\ 0 \\
  \vdots &\ \vdots &\ \ddots &\ \vdots\\
  0 &\ 0&\  \cdots &\ {H}^{\text{T}}(N^2)
  \end{array}%
\right],
\end{array}
\end{eqnarray*}
\begin{equation}
\begin{array}{c}
\mathcal{W}=\left[%
\begin{array}{cccc}
  F^{\text{T}}(1)  &\  {O}^{\text{T}}(1,2)  &\ \cdots  &\ {O}^{\text{T}}(1,N^2) \\
  {O}^{\text{T}}(2,1) &\  F^{\text{T}}(2) &\ \cdots   &\ {O}^{\text{T}}(2,N^2)) \\
  \vdots &\ \vdots &\ \ddots &\ \vdots\\
  {O}^{\text{T}}(N^2,1) &\ {O}^{\text{T}}(N^2,2)&\  \cdots &\
  F^{\text{T}}(N^2)\\
  \end{array}%
\right].\ \ \
\end{array}
\end{equation}
By making use of the  time-independent perturbation theory, we can
obtain high-order approximations  for Eq.(\ref{phieqn}) or
(\ref{phieqnmatrixform}). The zeroth order approximation solutions
%\begin{widetext}
\begin{eqnarray}
\begin{array}{c}
\widetilde{\Phi}_{\vec{k}_{\vec{r}}}^{[0]}(1)=\left[
\begin{array}{c}
  \Phi_{\vec{k}_{\vec{r}}}^{[0]}(1)  \\
  0 \\
  \vdots\\
  0\\
  \end{array}
\right]
\end{array},\begin{array}{c}
\widetilde{\Phi}_{\vec{k}_{\vec{r}}}^{[0]}(2)=\left[
\begin{array}{c}
  0 \\
  \Phi_{\vec{k}_{\vec{r}}}^{[0]}(2)  \nonumber\\
  \vdots\\
  0\\
  \end{array}
\right]
\end{array},\\
\cdots,\quad\cdots,\qquad\quad
\begin{array}{c}
\widetilde{\Phi}_{\vec{k}_{\vec{r}}}^{[0]}(N^2)=\left[
\begin{array}{c}
  0 \\
  0 \\
  \vdots\\
  \Phi_{\vec{k}_{\vec{r}}}^{[0]}{(N^2)} \\
  \end{array}
\right]
\end{array},
\end{eqnarray}
%\end{widetext}
are given by the eigenvalue equation
\begin{eqnarray}
{H}^{\text{T}}(n)\Phi_{\vec{k}_{\vec{r}}}^{[0]}(n)=E_{\vec{k}_{\vec{r}}}^{[0]}(n)\Phi_{\vec{k}_{\vec{r}}}^{[0]}(n)
\label{effspaceeqn}.
\end{eqnarray}
$\vec{k}_{\vec{r}}$ in the above equations represents a wave-vector
 defined by
$\vec{k}_{\vec{r}}=(\vec{k},i\vec{k}^{\text{A}})$.  This zeroth
order solution indicates  that in the six-dimension space, the
system and ancilla  move in an external potential $V(\vec{r})$ as a
whole system and evolve according  to the free Hamiltonian
$H^{\text{T}}(\vec{r})$. The interaction between the spin and the
magnetic field together with the spin relaxation  make the two
systems  feel a complex vector potential
$\vec{A}_{\vec{r}}^{\text{T}}$ which would in turn affect the
center-of-mass motion of the particle.

From these zeroth order solutions, we can obtain the first order
correction as
\begin{eqnarray}
&&E_{\vec{k}_{\vec{r}}}^{[1]}(n)=F^{\text{T}}(n)\nonumber\\
&&\widetilde{\Phi}_{\vec{k}_{\vec{r}}}^{[1]}(n){=}\sum_{
\stackrel{n'\neq n} {\vec{k}_{\vec{r}}\neq
\vec{k}_{\vec{r}}'}}\frac{\bra{\Phi_{\vec{k}_{\vec{r}}'}^{[0]}(n')}
{O}(n',n)\ket{\Phi_{\vec{k}_{\vec{r}}}^{[0]}(n)}}
{E_{\vec{k}_{\vec{r}}'}^{[0]}(n')-{E_{\vec{k}_{\vec{r}}}^{[0]}(n)}}
\widetilde{\Phi}_{\vec{k}_{\vec{r}}'}^{[0]}(n').~~~~
\end{eqnarray}
The condition with which we can safely neglect the first order
correction
 is
\begin{eqnarray}
\left|\frac{\bra{\Phi_{\vec{k}_{\vec{r}}'}^{[0]}(n')}
{O}(n',n)\ket{\Phi_{\vec{k}_{\vec{r}}}^{[0]}(n)}}
{E_{\vec{k}_{\vec{r}}'}^{[0]}(n')-{E_{\vec{k}_{\vec{r}}}^{[0]}(n)}}\right|\ll1,~\vec{k}_{\vec{r}}',n'\neq
\vec{k}_{\vec{r}},n.
\end{eqnarray}
This is the validity condition for the zeroth order BO
approximation.

In this section, we have extended the BO approximation to an open
system with spin relaxations. The zeroth order approximation was
defined as the regime where its effective Hamiltonian can be
diagonalized with fixed slowly changing variables. This
generalization is available for dissipations in all variables except
coordinates (or center-of-mass motion), which we shall study in the
next section.

\section{Dissipation of Center-of-Mass Motion\label{sec:coordinatedissipatin}}
We consider a moving particle coupled to  a boson bath, the
interaction between the particle and the bath can be described by
\cite{quantumnoisy}
\begin{eqnarray}
H_I={x}\otimes\sum_k g_k(b_k+b_k^{\dag}),
\end{eqnarray}
where ${x}$ is the coordinate operator for the particle, $b_k$ and
$b_k^{\dag}$ are the annihilation and creation operators for the
bath, respectively. By a standard procedure, we can derive a master
equation and write it in the form of Eq.(\ref{masterequation}), but
the dissipation term $\mathcal{L}$ in this case takes
\begin{eqnarray}
\mathcal{L}\rho&=&\gamma_1\left(2{x}\rho{x}-
\rho{x}^2-{x}^2\rho\right)\qquad\qquad\qquad\nonumber\\
&+&\gamma_2\left({x}{p}\rho+\rho{p}{x}-{x}\rho{p}-{p}\rho{x}\right).
\end{eqnarray}
Define $C=C^{\dag}=\sqrt{\gamma_1}{x}$ and
$D=D^{\dag}=\sqrt{\gamma_1}{x}-\frac{\gamma_2}{\sqrt{\gamma_1}}{p}$,
the dissipation term can be rewritten as $\mathcal{L}\rho=C\rho
D^{\dag}+D\rho
C^{\dag}-\frac12\{D^{\dag}C,\rho\}-\frac12\{C^{\dag}D,\rho\}$, which
can not be arranged in the Lindblad form \cite{Breuer2004}.
Following the same method  as that  in the last section, we obtain
an effective Hamiltonian  similar to Eq.(\ref{EH}), but in  this
section, we replace $L_S$ by $L_C$ with
$L_C=\frac{i}2[2\gamma_1{x}^2-\gamma_2({x}{p}+{p}{x})]-
\frac{i}2[2\gamma_1({x}^{\text{A}})^2-\gamma_2({x}^{\text{A}}{p}^{\text{A}}
+{p}^{\text{A}}{x}^{\text{A}})]
-i[2\gamma_1{x}{x}^{\text{A}}-\gamma_2({x}{p}^{\text{A}}+{p}{x}^{\text{A}})]$.
We choose the eigenfunctions of  $H_S^{\text{T}}=H_S-H_S^{\text{A}}$
as the bases, which are a direct product of the eigenfunctions of
$H_S$ and $H_S^{\text{A}}$. Let $H_S$ and $H_S^{\text{A}}$ have the
eigenfunctions $\ket{\chi_m(x)}$ and
$\ket{\chi_m^{\text{A}}(x^{\text{A}})}$ with the same eigenvalues
$\varepsilon_m$ for $x=x^{\text{A}}$, we can expand the eigenstate
of $\mathcal{H}_T$ by
$\{\ket{\Lambda_{mn}(x,x^{\text{A}})}=\ket{\chi_m(x)}\ket{\chi_n^{\text{A}}(x^{\text{A}})}\}$
as $ \ket{\widetilde{\Phi}}=\sum_{mn}\Phi_{mn}\ket{\Lambda_{mn}}$.
Following the same derivation  presented  in
Sec.\ref{sec:spindissipation}, we obtain the following equation  for
$\Phi_{mn}$ as,
\begin{eqnarray}
{H}^{\text{T}}(mn)\Phi_{mn}+F^{\text{T}}(mn)\Phi_{mn}+\sum_{pq\neq
mn}{O}^{\text{T}}(mn,pq)\Phi_{pq}\nonumber\\
+\sum_{pq\neq
mn}{L}_C^{\text{T}}(mn,pq)\Phi_{pq}=E\Phi_{mn}.~~~\label{phieqnDissCOM}
\end{eqnarray}
where
\begin{eqnarray}
H^{\text{T}}(mn)&=&-\frac1{2M}[\nabla_{\vec{r}}-i\vec{A}^{\text{T}}_{\vec{r}}(mn)]^2\nonumber\\
&+&V^{\text{T}}(\vec{r})+E_{mn}(\vec{r})+{L}_C^{\text{T}}(mn),\nonumber
\end{eqnarray}
and
\begin{eqnarray}
{L}_C^{\text{T}}(mn,pq)\Phi_{pq}=\bra{\Lambda_{mn}}L_C\ket{\Lambda_{pq}}.\nonumber
\end{eqnarray}
When the two subscripts in  ${L}_C^{\text{T}}$ are same, we will
omit one of them to shorten the notation. $pq\neq mn$ in the
summation means $(p, m)$ and  $(q, n)$ can not be taken equal
simultaneously. Other terms are similar to the results given in
Sec.\ref{sec:spindissipation} as
$F^{\text{T}}(n)=-\frac1{2M}\sum_{pq\neq
mn}(\inner{\Lambda_{mn}}{\nabla_{\vec{r}}\Lambda_{pq}}\inner{\Lambda_{pq}}{\nabla_{\vec{r}}\Lambda_{mn}}),
{O}^{\text{T}}(mn,pq)=-\frac1{2M}(2\inner{\Lambda_{mn}}{\nabla_{\vec{r}}\Lambda_{pq}}\nabla_{\vec{r}}+
\inner{\Lambda_{mn}}{\nabla^2_{\vec{r}}\Lambda_{pq}})$. We would
like to note that because of the bases we choose are uncoupled
between the quantum system and ancilla, the Hamiltonian
$H^{\text{T}}(mn)$ governing the center-of-mass motion  of the
system can be factorized as
$H^{\text{T}}(mn)=-\frac1{2M}[\nabla-i\vec{A}(m)]^2+V+\varepsilon_m-
(-\frac1{2M}[\nabla^{\text{A}}-i\vec{A}^{\text{A}}(n)]^2+V^{\text{A}}+\varepsilon_n)
+{L}_C^{\text{T}}(mn)$ with
$\vec{A}(m)=i\inner{\chi_m}{\nabla\chi_m}$ and
$\vec{A}^{\text{A}}(n)=i\inner{\chi_n^{\text{A}}}{\nabla^{\text{A}}\chi_n^{\text{A}}}$,
where $\nabla^{\text{A}}$ represents the derivative with respect to
$\vec{x}^{\text{A}}$. This means that the interaction between the
system and the magnetic field makes the system and the ancilla feel
vector potentials $\vec{A}(m)$ and $\vec{A}^{\text{A}}(n)$,
respectively. DissCOM induces a correction ${L}_C^{\text{T}}(mn)$ to
the system via  the interaction between the system and ancilla in
$H^{\text{T}}(mn)$. In the same way, namely treat the terms
$F^{\text{T}}$, $O^{\text{T}}$ and ${L}_C^{\text{T}}(mn,pq),(mn\neq
pq)$ as perturbations, we obtain  the validity condition  for the
zero order  BO approximation
\begin{eqnarray}
\left|\frac{\bra{\Phi_{\vec{k}_{\vec{r}}'}^{[0]}(pq)}
\left({O}^{\text{T}}(pq,mn)+{L}^{\text{T}}_C(pq,mn)\right)\ket{\Phi_{\vec{k}_{\vec{r}}}^{[0]}(mn)}}
{E_{\vec{k}'_{\vec{r}}}^{[0]}(pq)-{E_{\vec{k}_{\vec{r}}}^{[0]}(mn)}}\right|\ll1,\nonumber\\
mn,\vec{k}_{\vec{r}}\neq pq,
\vec{k}_{\vec{r}}',~~~\label{BOconditionDissCOM}
\end{eqnarray}
where
\begin{eqnarray*}
{L}_C^{\text{T}}(mn,pq)\Phi_{pq}=\bra{\Lambda_{mn}}L_C\ket{\Lambda_{pq}},
\end{eqnarray*}
and
\begin{eqnarray*}
L_C&=&\frac{i}2[2\gamma_1{x}^2-\gamma_2({x}{p}+{p}{x})]\nonumber\\
&-&\frac{i}2[2\gamma_1({x}^{\text{A}})^2-\gamma_2({x}^{\text{A}}{p}^{\text{A}}
+{p}^{\text{A}}{x}^{\text{A}})]\nonumber\\
& &-i[2\gamma_1{x}{x}^{\text{A}}
-\gamma_2({x}{p}^{\text{A}}+{p}{x}^{\text{A}})].
\end{eqnarray*}
From Eqs.(\ref{phieqnDissCOM}) and (\ref{BOconditionDissCOM}) we can
find that  the zero-order Hamiltonian $H^{\text{T}}$ only contains
the diagonal elements of ${L}_C^{\text{T}}$, while the  perturbation
matrix includes the spin part $O^{\text{T}}$, $F^{\text{T}}$, and
the spatially dependent term  ${L}_C^{\text{T}}(mn,pq),(mn\neq pq)$.
Recall that the main concept of BO approximation in our case  is to
separate  the spin variables from the spatial one (i.e., the fast
and slow variables). Due to the dissipation represented by
${L}_C^{\text{T}}$, this separation of those variables  becomes
difficult, resulting in the violation of BO approximation. We will
show in the next section that  this is different  for  the case of
spin relaxation, because the spin relaxation is coordinate
independent.

\section{Example\label{sec:example}}

\begin{figure}
\includegraphics*[width=0.8\columnwidth]{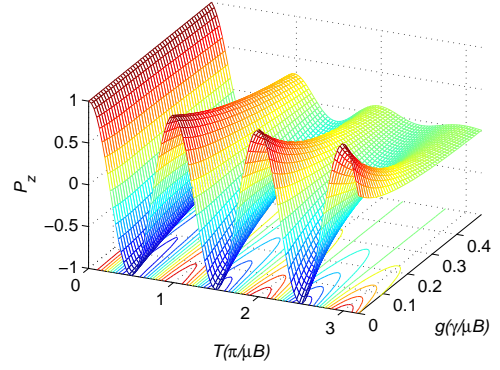}\caption{
The polarization of the neutron along the $z$ axis as functions of
the duration  $T$ and the rescaled dissipation rate $g$. The time
$T$ is plotted in units of $\pi/\mu B$. Initially the spin is in the
state $\ket{{+}\frac12}$, i.e. $P_z=1$. }\label{FIG:coupled}
\end{figure}

\begin{figure}
\includegraphics*[width=0.8\columnwidth]{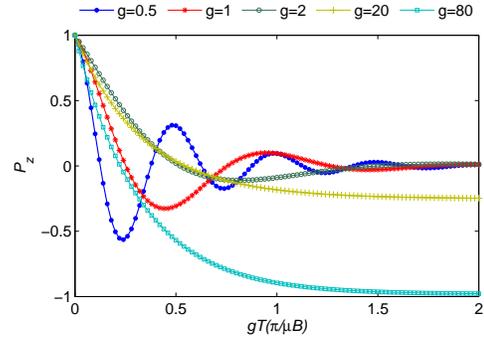}
\caption{(Color Online)The polarization of the neutron along the $z$
axis for different $g$ as a function of $gT$(in units of $\pi/\mu
B$).}\label{FIG:gfixed_T}
\end{figure}

As an example, in this section we consider a neutron moving in a
static helical magnetic field,
\begin{eqnarray}
\vec{B}=\vec{B}(z)=B\left(\sin\theta\cos\frac{2\pi z}L,
~\sin\theta\sin\frac{2\pi z}L,~\cos\theta\right).
\end{eqnarray}
The Hamiltonian for such a system is
\begin{eqnarray}
H=H(z)=\frac{{\vec{p}}^2}{2M}+\mu\vec{B}\cdot\vec{\sigma}={H}_K+H_S.
\end{eqnarray}
For fixed but arbitrary $z$, the interaction Hamiltonian $H_S$ has
the eigenfunctions
\begin{eqnarray}
\ket{\chi_1(z)}=\left(
\begin{array}{c}
  \cos\frac{\theta}{2}\exp(-i\frac{2\pi z}L)\\
  \sin\frac{\theta}{2}\\
  \end{array}
\right)\nonumber\\
\ket{\chi_2(z)}=\left(
\begin{array}{c}
  \sin\frac{\theta}{2}\exp(-i\frac{2\pi z}L)\\
  -\cos\frac{\theta}{2}\\
  \end{array}
\right),
\end{eqnarray}
and corresponding eigenvalues $\varepsilon_{1,2}=\pm\mu B$. This
system is the same as that studied  in Ref.\cite{CPSun1990}. Taken
the spin relaxation into account (see Eq.(\ref{spindissipation})),
the  effective Hamiltonian related to the spin reads
\begin{widetext}
\begin{eqnarray}
H_S^{\text{T}}(\varphi,\varphi^{\text{A}})=\mu B\left(
\begin{array}{cccc}
  -ig &\ -\sin\theta e^{i\varphi^{\text{A}}} &\ \sin\theta e^{-i\varphi} &\ 0\\
  -\sin\theta e^{-i\varphi^{\text{A}}} &\ 2\cos\theta-\frac12ig &\ 0 &\  \sin\theta e^{-i\varphi}  \\
  \sin\theta e^{i\varphi} &\ 0 &\ -2\cos\theta-\frac12ig &\ -\sin\theta e^{i\varphi^{\text{A}}}\\
  ig &\ \sin\theta e^{i\varphi} &\ -\sin\theta e^{-i\varphi^{\text{A}}} &\ 0\\
  \end{array}
\right),
\end{eqnarray}
\end{widetext}
where $\varphi=\frac{2\pi z}L$ and $\varphi^{\text{A}}=\frac{2\pi
z^{\text{A}}}L$. The rescaled coupling  constant is defined by
$g=\frac{\gamma}{\mu B}$ and therefore it is dimensionless. In the
following, we set $\theta=\frac{\pi}2$. The eigenvalue $E_j$ of
$H_S^{\text{T}}(\phi,\phi^{\text{A}})$ in this case are given by (in
units of $\mu B$)
\begin{eqnarray}
&&E_1=-\frac12gi,\nonumber\\
&&E_j^3+\frac32giE_j^2-\frac12(8+g^2)E_j+2gi(e^{-i(\varphi-\varphi^{\text{A}})}-1)=0,\nonumber\\
&&\qquad\qquad\qquad\qquad\qquad\qquad\qquad\qquad\quad(j=2,3,4),\nonumber
\end{eqnarray}
the corresponding right eigenstates are
\begin{eqnarray*}
\ket{R_j}=\left(
\begin{array}{c}
  A_j\\
  B_j\\
  C_j\\
  D_j\\
  \end{array}
  \right),
\end{eqnarray*}
while the left eigenstates read
\begin{eqnarray*}
\bra{L_j}=\frac1{N_j}\left(a_j,b_j,c_j,d_j\right).
\end{eqnarray*}
Here
\begin{eqnarray*}
%M_j=|A_j|^2+|B_j|^2+|C_j|^2+|D_j|^2,\\
&&N_j=(A_ja_j+B_jb_j+C_jc_j+D_jd_j).
\end{eqnarray*}
For $j=1$,
\begin{eqnarray*}
&&A_1=D_1=a_1=d_1=0,\\
&&B_1=e^{-i\varphi^{\text{A}}},~~C_1=e^{i\varphi},\\
&&b_1=e^{i\varphi^{\text{A}}},~~~~c_1=e^{-i\varphi},
\end{eqnarray*}
and for $j=2,3,4$,
\begin{eqnarray*}
&&A_j=4-giE_j-2E_j^2,\\
&&B_j=-2e^{-i\varphi}gi+2e^{-i\varphi^{\text{A}}}E_j,\\
&&C_j=2e^{i\varphi^{\text{A}}}gi-2e^{i\varphi}E_j,\\
&&D_j=4e^{i(\varphi-\varphi^{\text{A}})}+g(g-2iE_j),\\
&&a_j=e^{i(\varphi-\varphi^{\text{A}})}(4-giE_j-2E_j^2),\\
&&b_j=2e^{i\varphi}E_j, \\
&&c_j=-2e^{-i\varphi^{\text{A}}}E_j,\\
&&d_j=4.
\end{eqnarray*}
%\begin{eqnarray}
%\end{eqnarray}
We can see from the eigenvalues and eigenstates that only when
$\varphi=\varphi^{\text{A}}$, i.e. $z=z^{\text{A}}$, there exists a
 steady state for the quantum system
\cite{EHAHuang2008}, which is independent of initial state. This is
exactly the situation we will with in the following discussion, in
which we will discuss  the population transfer among the internal
states for the quantum system. Initially, the spin of the neutron is
prepared in the state $\ket{{+}\frac12}$, we manipulate the neutron
moved from $z=0$ to $z=L$ in a time interval $T$ and calculate  the
polarization of the neutron along the $z$ axis at time $T$. The
polarization of the neutron along $z$ axis at time $T$ with
different rescaled dissipation rate is plotted in
Fig.\ref{FIG:coupled}.  Several features can be found from these
figures. (1) When $g\rightarrow 0$, the relation between the
polarization along $z$ axis  and the duration  $T$ is a cosine
function. This coincides with the results given  in
Ref.\cite{CPSun1990}, meaning that our description can return back
to the results for closed systems. (2) As $g$ increases, the
relation between the polarization along $z$ axis and duration $T$ is
also an oscillating function, but the amplitude of the oscillation
decreases with the interval  $T$. (3) With $g\rightarrow0.5$ and
$T\rightarrow 3(\pi/\mu B)$, % (or $gT\rightarrow1.5(\pi/\mu B)$, which
%can be seen more strikingly in Fig.\ref{FIG:gfixed_T}),
the polarization along $z$ axis reaches zero, arrives at a steady
value. These features all result from the competition between the
spin-field coupling represented by $H_S$ and the dissipation
$\mathcal{L}$. When $T$ and $g$ are large enough, the accumulation
of the dissipation leads the spin of the neutron  into the steady
state, a result of the balance between the spin-field coupling $H_S$
and the dissipation. We plot the polarization along $z$ axis as a
function of $gT$ for different  $g$ in Fig.\ref{FIG:gfixed_T}. We
can see from the figure that when $g\rightarrow\infty$, $P_z$
sharply  drops into -1, i.e. the spin dissipates  into the state
$\ket{{-}\frac12}$. This can be easily understood as follows. When
$g\rightarrow\infty$, the spin-field coupling $H_S$ can be
neglected, namely the dissipation dominates the dynamics of the
spin, which makes the spin relax to its ground state
$\ket{{-}\frac12}$.


\begin{figure}
\includegraphics*[width=0.7\columnwidth]{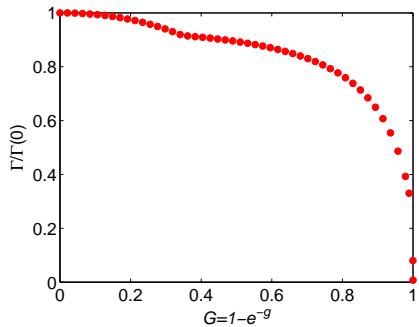}
\caption{Validity measure  $\Gamma(g)$ as a function of the rescaled
dissipation rate $g$. Other parameters in the figure are set to
satisfy $\frac1{\mu B M^2L}=10^{-6},~ \frac{k_z}{\mu
BML}=2\times10^{-4}$. The results have been normalized in units of
$\Gamma(g=0)$. }\label{FIG:BOconditioncheck}
\end{figure}

Finally, we discuss  the validity condition for the BO
approximation. For the example under discussion, if we do not
consider the dissipation, i.e., the system is closed, the validity
condition given in this paper returns to the validity condition
derived in  Ref.\cite{CPSun1990}, this means that the BO
approximation for open systems  defined in this paper can back to
closed systems. In order to see the effect of dissipation on the
validity condition,  we recall that without any dissipation, the
validity condition for this example is\cite{CPSun1990}
\begin{eqnarray*}
\frac1{\mu B M^2L}\ll1,~ \frac{k_z}{\mu BML}\ll1,
\end{eqnarray*}
where $k_z=\frac{v_z}M$. Namely, when the magnetic field is
homogeneous and strong enough as well as  the velocity of the
neutron along the $z$ axis is very small, the BO approximation is a
good approximation. In the following, we study how the dissipation
affects the validity condition for the open system. To show the
dependence of the validity condition on the dissipation rate, we
define the following function with $\max$ taken over all $m$ and $n$
except $m=n$,
\begin{eqnarray}
\Gamma(g){=}\max\left\{\left|\frac{\bra{\Phi_{\vec{k}_{\vec{r}}'}^{[0]}(n)}
{O}(n,m)\ket{\Phi_{\vec{k}_{\vec{r}}}^{[0]}(m)}}
{E_{\vec{k}_{\vec{r}}'}^{[0]}(n)-{E_{\vec{k}_{\vec{r}}}^{[0]}(m)}}\right|\right\},
\end{eqnarray}
to characterize the violation of the BO approximation. The numerical
results of $\Gamma(g)$ on $g$ is shown in
Fig.\ref{FIG:BOconditioncheck}.
In this figure, we have set all parameters except $g$ to satisfy the
BO approximation condition for closed system given in
Ref.\cite{CPSun1990} (e.g., $\frac1{\mu B M^2L}=10^{-6},~
\frac{k_z}{\mu BML}=2\times10^{-4}$), and have normalized the
results with respect  to $\Gamma(g=0)$, i.e. all the values in the
figure are rescaled  by $\Gamma(0)$. We can find that as $g$
increases, $\Gamma(g)$ decreases. This result tells us that if the
closed system satisfies the BO approximation condition, the
corresponding open system satisfies that condition too. Two points
are worth addressing.  (1) This conclusion depends sharply on the
dissipation,  for a given dissipation, we should check the
validation condition case by case, and (2) the dissipation benefits
the BO approximation in this example, this can be understood as
follows. In this example, the BO approximation for open systems is
defined   as the possibility to separate the spin and spatial
variables, the dissipation occurs only for the spin (slowly changed)
variable and it is coordinate independent, so it is reasonable that
the dissipation benefits the BO approximation. Similar conclusions
can be found in the adiabatic approximation in open systems, see
Ref.\cite{EHAYi2007,nMarkovShabani2005,EHAHuang2008}.

\section{Conclusion and Discussion\label{sec:CD}}
In this paper, we have extended BO approximation from closed  to
open systems by the effective Hamiltonian approach. Two types of
dissipation are considered,  the conditions under which the BO
approximation holds are given and discussed.  When the spin
relaxation/dissipation is considered, the center-of-mass of the
system  feels a complex vector potential depending on the
interaction between the spin and the magnetic field as well as the
dissipation rate. The zeroth order BO approximation was defined as
the regime in which its effective Hamiltonian can be diagonalized
with fixed slowly changing variables. When there is only DissCOM,
the vector potential can be factorized into two parts, which are for
the system and ancilla, respectively. The DissCOM introduces a
correction to the Hamiltonian for the center-of-mass motion and then
modifies the dynamics of the system. An example with spin
relaxation/dissipation has been presented and discussed.

\ \ \\
We thank Zheng-Yuan Xue for help. This work is supported by  NSF of
China under grant No. 10775023.

\end{document}